\documentclass[%
 reprint,
 amsmath,amssymb,
 aps,
]{revtex4-2}

\usepackage{graphicx}
\usepackage{dcolumn}
\usepackage{bm}
\usepackage{braket}
\usepackage{svg}
\usepackage{subcaption}
\usepackage{siunitx}
\usepackage{amsmath} 
\usepackage{tabularx}
\usepackage{makecell}
\usepackage{subcaption}
\usepackage{color}
\usepackage{booktabs}

\usepackage{hyperref}

\definecolor{darkblue}{rgb}{0.8, 0, 0}

\newcommand{\yso}[0]{Y$_2$SiO$_5$}

\newcommand{\euyso}[0]{$^{151}$Eu$^{3+}$:Y$_2$SiO$_5$}

\newcommand{\eu}[0]{Eu$^{3+}$}
\newcommand{\pr}[0]{Pr$^{3+}$}

\newcommand{\yttrium}[0]{Y$^{3+}$}

\newcommand{\eutransition}{$^7$F$_{0} \leftrightarrow ^5$D$_{0}$}
\newcommand{\gstate}{$^7$F$_{0}$}
\newcommand{\estate}{$^5$D$_{0}$}

\newcommand{\fone}{$\ket{5_g} \leftrightarrow \ket{6_e}$}   
\newcommand{\ftwo}{$\ket{3_g} \leftrightarrow \ket{6_e}$}   
\newcommand{\fthree}{$\ket{6_g} \leftrightarrow \ket{5_e}$}

\newcommand{\rftwo}{$\ket{3_g} \leftrightarrow \ket{5_g}$}

\newcommand{\norm}[1]{\left\lVert#1\right\rVert}

\begin{document}

\raggedbottom	

\title{Optical pumping simulations and optical Rabi frequency measurements in \euyso{} under magnetic field}

\author{Jingjing Chen}
\author{Mikael Afzelius}%
 \email{mikael.afzelius@unige.ch}
\affiliation{%
 Department of Applied Physics, University of Geneva, CH-1205 Geneva, Switzerland
}

\date{\today}

\begin{abstract}

Europium-doped crystal \yso{} is an interesting platform for optical quantum memories, due to its long optical and spin coherence times. In this work, we investigate \euyso{} under the application of magnetic field, a complex system with 36 different optical-hyperfine transitions. We present a simple numerical simulator for calculating the effect of optical pumping schemes that allow the isolation of a single frequency class of ions, applicable to any multi-level atom with inhomogeneous broadening. Experimentally, we develop methodologies for estimating the magnetic field vector, demonstrating high precision in predicting spectral features based on the known spin Hamiltonians. We measured the optical Rabi frequency of 21 transitions among the 36 possible optical-hyperfine transitions, allowing for the construction of a 6x6 branching ratio matrix of relative transition strengths. From the Rabi frequency measurements, we derive the optical dipole moment for the \eutransition{} transition, yielding a value of $(6.94 \pm 0.09) \times 10^{-33}~\mathrm{C} \cdot \mathrm{m}$. Our results provide a critical test of the spin Hamiltonian models, demonstrating high precision in predicting energy levels and relative transition strengths, which are key abilities for quantum applications with \euyso{} under magnetic field.
\end{abstract}


\maketitle

\section{Introduction}
\label{sec:intro}

Optical quantum memories \cite{Bussieres2013,Heshami2016} play an important role in quantum networks, potentially allowing long-distance quantum communication using quantum repeaters \cite{Briegel1998,Duan2001,Sangouard2011}. In this context, rare-earth (RE) doped crystals are highly interesting for ensemble-based quantum memories \cite{Simon2007,Sinclair2014}, particularly due to their long optical and spin coherence times, and large multimode storage capacity, see Ref. \cite{Tittel2010b} for a recent review.

Among RE-doped crystals, the non-Kramers ions \pr{} and \eu{} have been used in a range of experiments showing spin-wave storage where hyperfine states were exploited for on-demand recall and long-duration storage, both for classical pulses \cite{Ma2021,Heinze2013} and for quantum states, including storage of qubits \cite{Gundogan2015,Ortu2022b}, quantum correlations \cite{Laplane2017,Kutluer2017}, and entanglement \cite{Ferguson2016,Kutluer2019,Rakonjac2021}. These ions have a quenched electronic spin $S=0$ in most host crystals, such that their hyperfine structure is given by their nuclear spin $I=5/2$. At zero magnetic field, each electronic state has three doubly degenerate hyperfine levels due to a quadrupole-type interaction. With a magnetic field, each degenerate doublet splits due to the Zeeman effect, resulting in 6 hyperfine levels per electronic state, see eg. \cite{Longdell2002,Longdell2006}.

Many storage experiments in \pr{} and \eu{} have been realized at zero magnetic field, mostly due to convenience and the low number of hyperfine states. However, it has been observed that applying even a very weak magnetic field between 0.1 and 1 mT can substantially increase the spin coherence time \cite{Alexander2007,Ortu2022b}, as recently studied theoretically \cite{Pignol2024}. But, at these fields the Zeeman doublets are not resolved in optical quantum memory experiments, leading to oscillations in the storage efficiency as a function of spin storage time \cite{Etesse2021}. On the other hand, one can apply large fields to completely resolve all hyperfine transitions, which allows quenching efficiency oscillations \cite{Holzaepfel2020,Chen2025} and potentially reaching much longer spin coherence times if a Zero First-Order Zeeman (ZEFOZ) transition can be obtained \cite{Fraval2004,Longdell2005,Heinze2013,Ma2021,Zhong2015,Wang2025}. But this brings on new challenges, as the frequencies and strengths of the 36 possible optical transitions between the hyperfine levels (optical-hyperfine transitions) in \pr{} and \eu{} depend strongly on the magnetic field orientation and magnitude, due to the anisotropy of the doping site in the host crystal. Some of these challenges are addressed in this article for the case of \euyso{}.

Recently, we demonstrated that operating an atomic frequency comb memory (AFC) in \euyso{} at a magnetic field of about 230 mT, allowed us to reach a highly efficient and reversible conversion from the optical state to the spin state, reaching up to 96\% efficiency \cite{Chen2025}. In this article, we describe the key experimental methods that allowed us to operate the AFC spin-wave storage experiment at such a large field. These include methods for accurately tuning the magnetic field with respect to the symmetry axis of the \yso{} crystal, and the accurate estimation of the resulting B field vector in the reference frame of the crystal, using a combination of Raman heterodyne scattering (RHS) \cite{Mlynek1983} and spectral hole burning (SHB) \cite{MacfarlaneShelby1987,LuiJacquierBook2025} techniques.

A key tool in these methods is the comparison of measured RHS and SHB frequencies with those predicted by the spin Hamiltonians \cite{Longdell2006,ZambriniCruzeiro2018a,Ma2018}, both for the optical ground and excited states. Indeed, accurate spin Hamiltonians are critical to reliably predict the hyperfine energy manifolds, but also for estimating the relative transition strength of the 36 optical-hyperfine transitions, i.e. the 6x6 branching ratio table between the ground and excited hyperfine levels. The branching ratio table, in turn, is key to a proper selection of the frequencies involved in the optical pumping and operation of the AFC quantum memory, eg. Ref. \cite{Chen2025}.

A severe test of the spin Hamiltonians is to compare the branching table elements to measured transition strengths, as these depend on delicate choices of angles and signs of the tensor elements of the spin Hamiltonians. These choices result in energy manifolds that are very difficult to differentiate with only energy measurements, as discussed in Ref. \cite{ZambriniCruzeiro2018a}, but they result in different branching tables. To this end, we measured the Rabi frequency of 21 transitions among the 36 optical-hyperfine transitions. With these data in hand, the 6x6 table of relative transition strengths could be estimated directly from the experiment. The experimental table is in excellent agreement with the one predicted by the choice of spin Hamiltonians given in Ref. \cite{ZambriniCruzeiro2018a}. In addition, the results allowed us to estimate the absolute transition strenght, i.e. the optical dipole moment, of the \eutransition{} transition in \euyso{}.

This article is organized as follows. In Section~\ref{sec:qm_with_field} we present the effective Hamiltonian describing the magnetic properties of hyperfine levels in \euyso{}. In Section~\ref{sec:simulation_opt_pump} we discuss optical pumping simulations used to isolate individual frequency classes, along with the generalized numerical model we employ. This model, based on population redistribution via optical excitation and spontaneous emission, can be adapted to other pumping schemes and inhomogeneously broadened multi-level systems. In Section~\ref{sec:experiment_setup}, we describe the experimental setup. Section~V shows the methods used to estimate the magnetic field vector using RHS and SHB techniques. In Section~\ref{sec:branching_ratio_estimation}, we present the measurement of optical Rabi frequencies and the fitting of the branching ratio table. We conclude in Section~\ref{sec:discussion} with a discussion of the implications of our findings and an outlook for future work.

\section{Quantum memory in $^{151}$Eu$^{3+}$:Y$_2$SiO$_5$ with lifted Zeeman degeneracy}
\label{sec:qm_with_field}

\subsection{Europium-doped \yso{}}
\label{sec:qm_with_field_A}

The \yso{} host crystal has a monoclinic structure and belongs to the $C_{2h}^6$ space group. Optically, it is a biaxial crystal, where the crystallographic $b$ axis is one of the principal axes of the dielectric tensor. The other two dielectric axes lie in the $a$-$c$ plane and are labelled $D_1$ and $D_2$ \cite{Li1992}. For optical experiments, crystals are conventiently cut along the orthogonal $b$, $D_1$, and $D_2$ axes, and spin tensors are often expressed in this reference frame.

In \yso{}, the \eu{} ions substitute for \yttrium{} ions at one of two crystallographic sites, denoted sites 1 and 2, both possessing a low $C_1$ point symmetry, where we use ions in site 1. Europium has two stable isotopes, Eu-151 and Eu-153, at roughly the same abundance. In this work, we use a crystal that has been isotopically enriched with Eu-151, from the same crystal boule that provided the crystals used in Refs. \cite{Jobez2015,Laplane2017,Ortu2022b}. Note that the isotopic frequency shift is smaller than the optical inhomogeneous broadening. The \eu{} ions are excited on the \eutransition{} transition, where \gstate{} is the ground state and \estate{} the excited state, see Fig. \ref{fig:energy_diagram}. For site 1, the transition frequency is 17240.2 cm$^{-1}$ (580.04 nm in vacuum).

\subsection{The spin Hamiltonian}
\label{sec:qm_with_field_B}

In a low-symmetric doping site, \eu{} has no net electronic spin $S=0$, and both isotopes have nuclear spin $I = \frac{5}{2}$. The hyperfine manifolds of each of the electronic states can be described by an effective spin Hamiltonian. Here we work with the usual form \cite{Longdell2006}

\begin{equation}
   H = \mathbf{I} \cdot \mathbf{Q} \cdot \mathbf{I}  + \mathbf{B} \cdot \mathbf{M} \cdot \mathbf{I}
\end{equation}

\begin{figure}
    \centering
    \includegraphics[width=\linewidth]{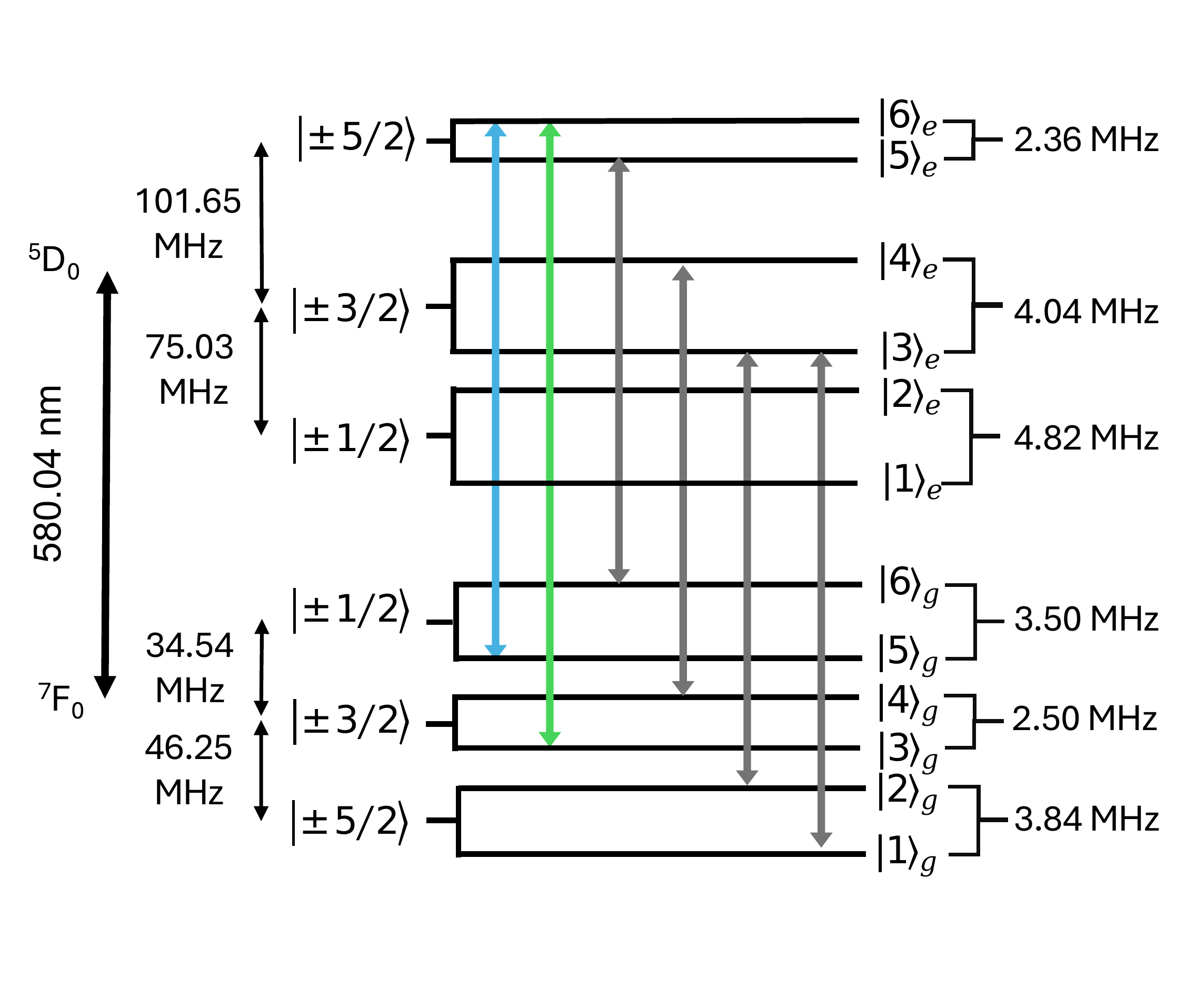}
    \caption{Hyperfine structure of the electronic states of \euyso\ under an external magnetic field $B$ of 230 mT along the $D_2$ axis. The levels are labeled as $\ket{1_g}$ to $\ket{6_g}$ for the $^7F_0$ ground states, and $\ket{1_e}$ to $\ket{6_e}$ for the $^5D_0$ excited states. We also show the set of frequencies used in the AFC spin-wave experiment in Ref. \cite{Chen2025}, in the present RHS experiments and some of the optical Rabi frequency measurements. The AFC scheme involves an input/output transition (blue) and an optical control transition (green). Optical pumping transitions (gray), used in the RHS measurement for class cleaning and population initialization into the $\ket{5_g}$ state, are also indicated.}
    \label{fig:energy_diagram}
\end{figure}

\noindent Here, $\mathbf{B}$ represents the external magnetic field vector, while $\mathbf{M}$ and $\mathbf{Q}$ are the effective Zeeman and quadrupole tensors, and $\mathbf{I}$ is the nuclear spin-operator vector. The quadrupole term, which includes pure quadrupolar and pseudoquadrupolar contributions, causes a split of the nuclear states of the order of 10 to 100 MHz for europium isotopes in \yso{}, forming three degenerate doublets at zero magnetic field. For 151-Eu in \yso{} specifically, the zero-field quadrupolar splits are 34.54 and 46.25 MHz for the ground state \gstate{} \cite{Longdell2006,ZambriniCruzeiro2018a}, and 75.03 and 101.65 MHz \cite{Ma2018} for the excited state \estate{}, see Fig. \ref{fig:energy_diagram}. Applying a magnetic field lifts the degeneracy due to the Zeeman term, with a Zeeman effect of the order of 1-10 MHz/T. For the fields in the range of $230-250$ mT used in this work, the quadrupolar splits dominate the hyperfine structure, and the weak Zeeman term splits are in the range of 1-4 MHz for our field orientation.

The properties of the quadrupole $\mathbf{Q}$ and Zeeman $\mathbf{M}$ tensors are described elsewhere \cite{Longdell2006,ZambriniCruzeiro2018a,Ma2018}. Here we recall that for the low $C_1$ point symmetry, the tensor directions are arbitrary and have three different principal values. Numerical diagonalization of the spin Hamiltonians leads to two sets of 6 nuclear eigenstates, $\ket{i_g}$ and $\ket{j_e}$ for the ground and excited states, respectively, with integers $i,j = [1,6]$. We assume these are ordered from lowest to highest values, for each electronic state. Under the assumption that the electronic and nuclear wavefunctions are separable \cite{Mitsunaga1985,Bartholomew2016}, it follows that the optical dipole moment of transition $i_g \leftrightarrow j_e$ is $\mu_{i_g,j_e} = \mu_{g,e} \braket{j_e|i_g}$. Here $\mu_{g,e}$ denotes the optical transition dipole moment between the ground \gstate{} and excited \estate{} state.
The relative probability of the transition is given by $\gamma_{i,j} = |\braket{j_e|i_g}|^2$, and the set of all relative transition probabilities forms a 6x6 matrix of branching values.

For each crystallographic site, there are two non-equivalent magnetic subsites, which are related by a $C_2$ symmetry axis along the $b$ axis. This means that they have identical principal values for the $\mathbf{Q}$ and $\mathbf{M}$ tensors, but different orientations, e.g. the site 2 $\mathbf{Q}$ tensor can be generated from the site 1 $\mathbf{Q}$ tensor by applying the $C_2$ rotation around the $b$ axis. For a magnetic field that is neither parallel nor perpendicular to the $b$ axis, the two magnetic subsites have different hyperfine eigen energies and states. They thus act as two distinct ensembles, and only one can be used at any time for a quantum memory experiment. This reduces the available optical depth, and hence the memory efficiency. However, by aligning the field along the $b$ axis, or in the $D_1$-$D_2$ plane, the two sub-ensembles become equivalent, with the same hyperfine splittings and relative transition moments. Working in the $D_1$-$D_2$ plane generally allows tuning the hyperfine properties, as any angle in the plane results in equivalent magnetic subsites. In this work, we align the field close to the $D_2$ axis, as motivated in Sec. \ref{sec:simulation_opt_pump_EuYSO_example}.

\section{Optical pumping simulations in inhomogeneously broadened systems}
\label{sec:simulation_opt_pump}
\subsection{Optical pumping for isolation of a single frequency class of ions}
\label{sec:simulation_opt_pump_intro}

In rare-earth ion doped crystals, for non-Kramers ions it is usually the case that the optical inhomogeneous broadening is much larger than the spectrum covered by all the optical-hyperfine transitions. In this case, the spectrum of a single ion is completely hidden within the inhomogeneous broadening. For an ion with $N_g$ ($N_e$) ground (excited) hyperfine states, a monochromatic laser excites $N_g \times N_e$ transitions simulataneously. They belong to different frequency classes, each representing ions with a different frequency detuning $\delta_{g,e}$ within the inhomogeneous broadening. For \euyso{} with an applied field, there are 36 different frequency classes.

To be able to address specific transitions, one can apply optical pumping sequences in order to "distill" a single frequency class \cite{Nilsson2004,Lauritzen2012}, such that at specific frequencies within the broadening, one is sure to interact with a specific transition of a single frequency class. We refer to this step in the optical pumping as class cleaning. A simple and quite general recipe consists in selecting $N_g$ different optical-hyperfine transitions among the $N_g \times N_e$ possible transitions, such that all the ground states are excited for a single frequency class. Applying these frequencies, for any arbitrary relative shift of the frequencies within the inhomogeneous broadening, will often result in all other frequency classes being pumped away at these frequencies, provided that efficient optical pumping is possible, leaving only the selected frequency class at these particular frequencies. Scanning the pump frequencies over some range around these frequencies creates "trenches" of absorption where now we have only a single frequency class, such that in each trench we are certain which transition is excited. This is clearly important for quantum memory applications, but is also very useful for measuring, for instance, the optical Rabi frequency for a specific optical-hyperfine transition \cite{Lauritzen2012}. An important question for applications is the maximum width of the trenches, which sets an upper bound of the quantum memory bandwidth. To study different sets of class cleaning frequencies, it is useful to simulate the effect of the pump frequencies on the inhomogeneous spectrum.

\subsection{Numerical model}
\label{sec:simulation_opt_pump_num_model}

\begin{figure*}
    \centering
    \includegraphics[width=\linewidth]{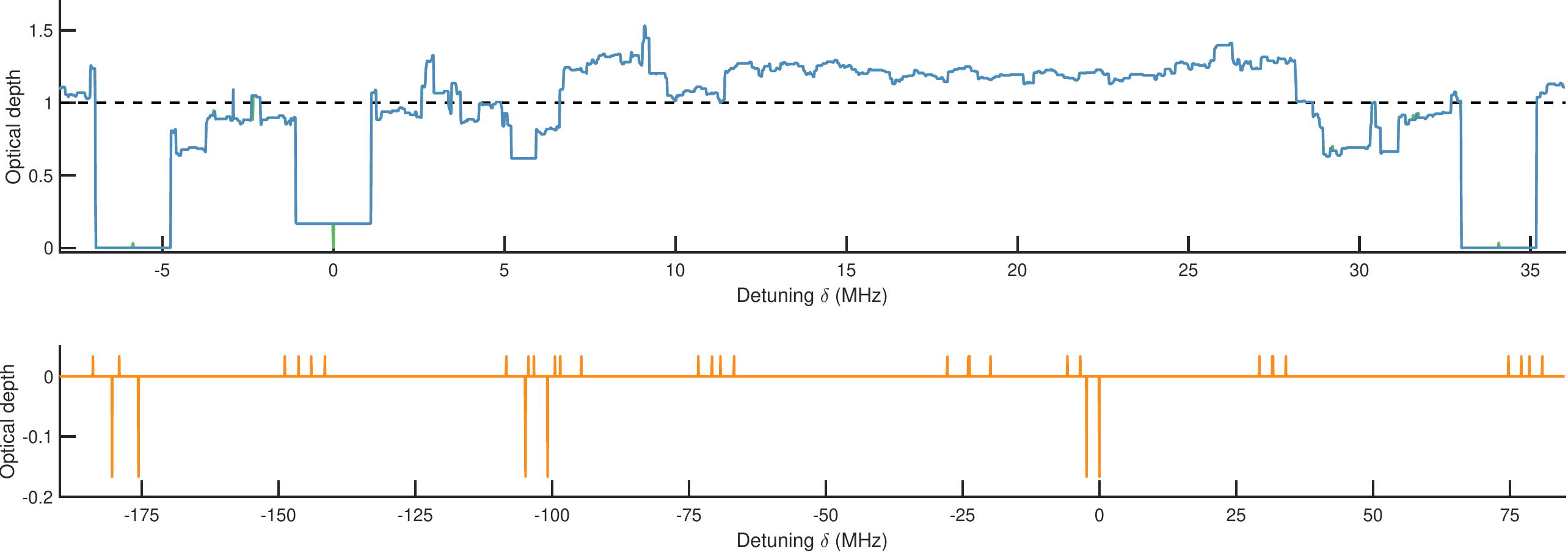}
    \caption{\textbf{Numerical simulation of the spectrum.} (a) The blue curve shows the simulated optical depth after the CC and SP steps, using the six CC frequencies shown in Fig. \ref{fig:energy_diagram}, for a field of 230 mT along the $D_2$ axis. The \fone{} transition was set to the relative frequency of zero, i.e. it is in the centre of the inhomogeneous broadening. Each frequency was swept over 2.2 MHz, which creates trenches in the absorption spectrum, as seen clearly at the zero-frequency \fone{} transition. Two additional CC trenches can be seen, at the relative transition frequencies -5.9 and 34.1 MHz, corresponding to the \fthree{} and \ftwo{} CC transitions. The green curve shows the spectrum after doing SHB on the \fone{} transition. The dashed black line shows the unpumped spectrum, which has been normalized to one, and all other spectra are normalized to the unpumped spectrum. (b) Subtracting the CC+SP spectrum (blue curve in (a)) from the CC+SP+SHB spectrum (green curve in (a)) highlights the holes and anti-holes due to the SHB, corresponding to negative and positive relative optical depths, respectively. One observes 6 holes and 30 anti-holes, as expected from a single frequency class.}
    \label{fig:spectrum_simulation}
\end{figure*}

Here we will describe a simple numerical model for simulating the effects of a sequence of pump pulses at different frequencies, including frequency-chirped pulses, in an inhomogeneously broadened system. We suppose that the hyperfine energies of the ground and excited states are $g_i$ and $e_j$, respectively, where the lowest energy states are $g_1=0$ and $e_1=0$. The transition $\ket{i_g} \leftrightarrow \ket{j_e}$ has frequency 

\begin{equation}
    T_{i,j}(\delta_{g,e}) = e_j - g_i + \delta_{g,e},
\end{equation}

\noindent where $\delta_{g,e}$ is the optical detuning parameter within the optical inhomogeneous broadening, which defines the frequency class. Note that in the following the $i$ index refers to the ground state, and the $j$ index to the excited state, and we often omit the $g$ and $e$ subscripts to simplify the notation, when appropiate.

The broadening is described by a probability distribution function $G(\delta_{g,e})$. In reality, this broadening is centered on the optical transition frequency, but for simulations it is convenient to center it at around $\delta_{g,e}=0$. The state of the ions is described by the density element $\rho_{i}(\delta_{g,e})$, which is the relative population in the ground state $\ket{i_g}$ for the ion belonging to the frequency class $\delta_{g,e}$, such that

\begin{equation}
    \sum_{i = 1} ^{N_g} \rho_{i}(\delta_{g,e}) = 1.
\end{equation}

\noindent In our case, the initial thermal population is equally distributed over all 6 states, that is $\rho_{i}(\delta_{g,e}) = 1/6$.

The sequence of pump pulses is numerically described by a discrete vector of pump frequencies $\delta_{pump} = [\delta_1,\delta_2, \delta_3, ...]$. For chirped pulses, the frequency chirp is entered into the vector using the same frequency step as in the optical detuning vector $\delta_{g,e}$. The vector $\delta_{pump}$ thus contains all frequency components of the sequence of pulses, centered at different frequencies, including possible repetition of the pulses.

In a first step we compute the ion state change due to the pump pulses. The code loops over the elements $k$ in the $\delta_{pump}$ vector. For each pump frequency $\delta_{pump}(k)$, we find an ion that fulfills the resonant condition

\begin{equation}
    T_{i,j}(\delta_{g,e}) - \delta_{pump}(k) = 0
\end{equation}

\noindent That is, we find a frequency class $\delta_{g,e}$ that resonates with the pump frequency $\delta_{pump}(k)$, on the transition $\ket{i_g} \leftrightarrow \ket{j_e}$. For this class, we take the population of the pumped ground state $\rho_{i}(\delta_{g,e})$ and we distribute this population over all the ground states of class $\delta_{g,e}$, according to the optical branching table elements $\gamma_{i,j}$. This simulates a situation where, for each resonance, we excite all the ions in state $\ket{i_g}$ to the excited state $\ket{j_e}$, and we wait enough time for all the population to decay back to the ground states, due to spontaneous emission. A more elaborate model could describe the excitation process more carefully, taking into account the power at each frequency and the transition strength. Note that, for a large inhomogeneous broadening as compared to the range covered by the transitions $T_{i,j}$, there will be $N_g \times N_e$ frequency classes resonating with any given pump frequency $\delta_{pump}(k)$, each with a different detuning $\delta_{g,e}$ (i.e. different frequency classes). The goal with the class cleaning is to find a set of frequencies where only a single frequency class resonates simultaneously with all the $N_g$ class cleaning transitions, such that all other frequency classes are pumped away into hyperfine states that do not resonate with these frequencies, effectively making them transparent.

In a second step, we compute the optical inhomogeneous absorption spectrum by assuming that each ion is described by a homogenous lineshape function $H(\delta)$, which is typically a Lorentzian function. The spectrum is given by

\begin{equation}
    f(\delta) = \chi \sum_{\delta_{g,e}} G(\delta_{g,e}) \left[  \sum_{i,j} \gamma_{i,j} \rho_{i}(\delta_{g,e}) H(\delta - T_{i,j}(\delta_{g,e}) ) \right]
    \label{eq:theory_spectrum}
\end{equation}

\noindent where $\chi$ is a normalisation parameter to obtain the correct absorption depth after the summation. Essentially, in the bracket of Eq. (\ref{eq:theory_spectrum})  we compute the line spectrum for each frequency class, which in our case consists of 36 optical-hyperfine lines for each class, where we weigh each line with its strength and the population in the corresponding ground state. Then we sum up all the line spectra in the entire inhomogeneous broadning. The calculation effectively represents the convolution of the single-ion line spectrum with the inhomogeneous distribution function $G(\delta_{g,e})$.

\subsection{Simulation example for 230 mT along $D_2$ in \euyso{}}
\label{sec:simulation_opt_pump_EuYSO_example}

Here we provide an example of a simulation using the 6 class cleaning frequencies that we employed in Ref. \cite{Chen2025}, for a simulated magnetic field of 230 mT along the $D_2$ axis. The frequencies are also shown in Fig. \ref{fig:energy_diagram}. Among these transitions, the $\ket{5_g} \leftrightarrow \ket{6_e}$ and $\ket{3_g} \leftrightarrow \ket{6_e}$ transitions formed the $\Lambda$-system in the AFC spin-wave optical storage experiment, see Ref. \cite{Chen2025} for details. In a first step, all 6 class cleaning (CC) frequencies are applied to distill a single frequency class, and in a second step all transition except the $\ket{5_g} \leftrightarrow \ket{6_e}$ were applied in order to spin polarize (SP) all ions into the $\ket{5_g}$ state. This maximizes the optical depth on the $\ket{5_g} \leftrightarrow \ket{6_e}$ transition, which was used to absorb the input pulse in the storage experiment. For other spectroscopic measurements, eg. the optical Rabi frequency measurements in this work, the ions of the single class can be polarized into any of the other states.

\begin{figure}
    \centering
    \includegraphics[width=\linewidth]{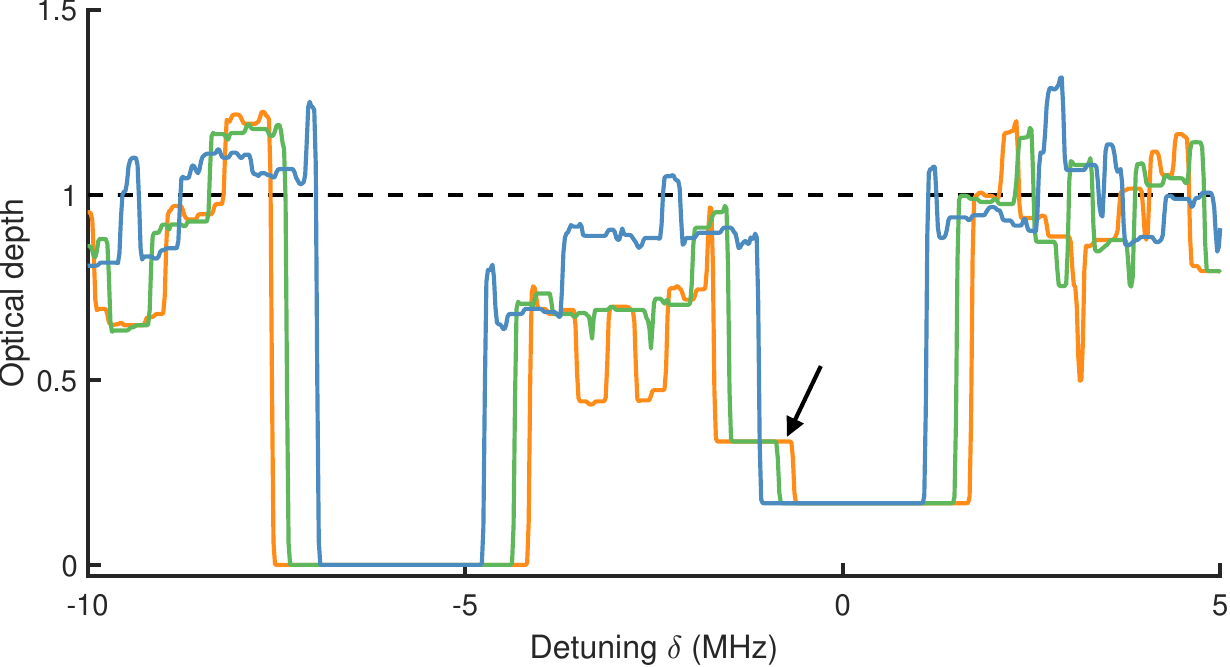}
    \caption{\textbf{Bandwidth limitation.} Numerical simulation of the CC and SP steps, using the 6 frequencies shown in Fig. \ref{fig:energy_diagram}. The curves show the absorption spectrum in the zero-frequency trench, centered on the \fone{} transition of the selected single-frequency class, for increasing bandwidth of the CC and SP pulses. The bandwidths were 2.2 MHz (blue curve), 3 MHz (green curve) and 3.4 MHz (orange curve). For bandwidths larger than the $\ket{5_e} - \ket{6_e}$ energy separation of 2.36 MHz, the trench displays a step, as indicated by the black arrow, which is due to some additional frequency class contributing to the absorption spectrum. Setting the bandwidth to the $\ket{5_e} - \ket{6_e}$ energy separation generally produces flat absorption trenches of a single class.}
    \label{fig:spectrum_simulation_pitwidth}
\end{figure}

In Fig. \ref{fig:spectrum_simulation}(a), we show the result of the simulation after the CC and SP steps, compared to the unpumped spectrum. In addition, we show the result of spectral hole burning (SHB) on the $\ket{5_g} \leftrightarrow \ket{6_e}$ transition, after the CC and SP steps, which shows a narrow transmission hole and associated narrow anti-holes at the expected frequency separation from the burned hole \cite{MacfarlaneShelby1987}. To reveal the complete structure of holes and anti-holes, we subtract the CC+SP spectrum from the CC+SP+SHB spectrum, which then shows only the relative change in optical depth induced by the SHB on the single frequency class, as shown in Fig. \ref{fig:spectrum_simulation}(b). One can see in total 6 spectral holes, and 30 anti-holes, the exact number expected for a single class, and they appear at the expected frequencies for this frequency class. Note that two doublets of anti-holes at -23.8 and 31.6 MHz are unresolved in the plot. The simulation shows beyond doubt that the chosen CC frequencies do indeed produce a single frequency class, which is one of the main utilities of the simulator. It is easy to test other sets of CC frequencies, in order to access specific transitions of a single frequency class, which is a feature we used extensively for the optical Rabi frequencies as discussed in Sec. \ref{sec:Rabi_measurements}. Note that for the simulation shown in Fig. \ref{fig:spectrum_simulation}, the branching values were identical and set to $\gamma_{i,j} = 1/6$, which makes all the holes and anti-holes clearly visible, of equal strength, and easy to count. In the appendix \ref{app:simulation_true_dm}, we show the same simulation with the branching values as given by the spin Hamiltonian, which shows the absorption spectrum one would expect from an actual experiment.

The simulator can also be used to estimate the bandwidth over which one can prepare a single frequency class. The method consists in increasing the frequency bandwidth of the pulses, until one observes steps and discontinuities in the absorption profile inside the trenches, which ideally should be flat, see Fig. \ref{fig:spectrum_simulation_pitwidth}. One can also do the same SHB hole burning test as discussed above, and count the number of holes and anti-holes for different bandwidths, at different frequencies inside the trenches. This permits a more quantitive analysis of the number of classes contributing to the absorption in the trenches. This analysis reveals that in our case, the trench bandwidth is limited by the $\ket{5_e} - \ket{6_e}$ Zeeman split, which is 2.36 MHz for a field of 230 mT along the $D_2$ axis, see also Fig. \ref{fig:spectrum_simulation_pitwidth}. We also observed that this bandwidth is maximum close to the $D_2$ axis. This is the main motivation for choosing a field angle close to $D_2$ for the optical storage experiments in Ref. \cite{Chen2025}.

\section{Experimental set-up}
\label{sec:experiment_setup}

The \euyso{} crystal has dimensions $2.5 \times 2.9 \times 12.3 \times$ along the $D_1 \times D_2 \times b$ axes. It is glued with silverpaste to a mount that rests on a soft copper cushion, to reduce vibrations from the pulse tube cooler. In addition, the trigger of the experimental sequence is provided by a piezo attached to one of the He tubes of the pulse tube compressor, which allows us to find a relatively stable time window for high-resolution SHB measurements, within the 710 ms cryo cycle.

The magnetic field is provided by positioning the crystal between two NdFeB permanent magnets. Two different spacings were used in this work, giving roughly 230 and 250 mT, respectively. The crystal $D_2$ axis is aligned as close as possible to the field direction, to within about 10 degress. To accurately tune the field into the $D_1-D_2$ plane, we use a pair of Helmholtz coils placed outside the cryostat, aligned as close as possible along the $b$ axis. Each coil pair is independently driven by a programmable current controlled power supply that has a calibrated conversion factor of 0.9119 mT/A and 0.7602 mT/A respectively, allowing for fine-tuning of the compensation field magnitude. With this external field, we can cancel the small $b$-axis component of the field generated by the permanent magnets, as explained in Sec. \ref{sec:field_tuning_D1D2}. Typically this required a field of 10-15 mT from the coils. The resulting field then lies in the $D_1-D_2$ plane, with a small angle to the $D_2$ axis, which depends on the relative positions of the permanent magnets, crystal and Helmholtz coils.

For the RHS measurements, we excite the four radio-frequency (RF) transitions between the $\ket{3_g}$,$\ket{4_g}$ doublet and the $\ket{5_g}$,$\ket{6_g}$ doublet, in the range of 31 to 38 MHz. To ensure RF field homogeneity across the entire crystal length, we use a solenoid around the crystal with an axial length of 22 mm. To enhance the power transmission to the RF solenoid, we use the tunable resonant RF circuit proposed in Ref. \cite{Ortu2021a}. This circuit uses a combination of serial and parallel capacitors, placed both inside and outside the cryostat, to reach a 50-Ohm impedance match for a long solenoid with an inductance of about 1 $\mu$H. The circuit resonance has a line width of about 300 kHz, corresponding to a quality factor of roughly 110, which gave a Rabi frequency of 42 kHz on the \rftwo{} hyperfine transition.

The optical setup is identical to the one presented in Ref. \cite{Chen2025}, here we point out some its key features. The experiment involved three different optical beams that propagated co-linearly through the crystal. The optical preparation steps, i.e. the CC, SP and SHB sequences, were done using a mode with a large beam waist radius of 350 $\mu$m inside the crystal. The coherent driving required for the optical Rabi measurements were done with a mode having a beam waist of 102 $\mu$m. Readout of the absorption spectrum was done with a mode having a beam waist of 34 $\mu$m.

\section{Experimental results}

\subsection{Tuning the magnetic field into the D$_1$-D$_2$ plane}
\label{sec:field_tuning_D1D2}

In this section, we explain how we cancelled the small component of the magnetic field along the $b$ axis, to make sure that the resulting field lies in the D$_1$-D$_2$ plane. As explained in Sec. \ref{sec:qm_with_field_B} this simplifies the absorption spectrum and increases the optical depth, as the two magnetic sub sites become equivalent.

To this end, we measured the spin transitions between the $\ket{3_g}$, $\ket{4_g}$ and $\ket{5_g}$, $\ket{6_g}$ hyperfine states. For a single magnetic subsite, there are then four RF transitions $\omega_{3,5}$, $\omega_{3,6}$, $\omega_{4,5}$ and $\omega_{4,6}$, respectively. For a field outside the D$_1$-D$_2$ plane, one observes eight RF transitions, due to the two non-equivalent subsites. By carefully tuning the $b$-axis compensation field, the two sets of RF transitions can be merged into a single set of four transitions, which is an unambiguous signature of having the field in the $D_1$-$D_2$ plane.

The RF transitions were measured using the RHS technique \cite{Mlynek1983}. The process begins with a CC and SP step to polarize the population into a selected ground-state hyperfine level, as explained in Sec. \ref{sec:simulation_opt_pump_intro}. A weak probe beam is applied, while a RF field is simultaneously swept across the relevant hyperfine splittings within the ground-state manifold. When the RF field is in resonance with the spin transition, a coherent RHS signal is generated. The spin resonance is detected as a beat signal between probe beam and the RHS signal, at the corresponding spin frequency.

\begin{figure}
    \centering
    \includegraphics[width=\linewidth]{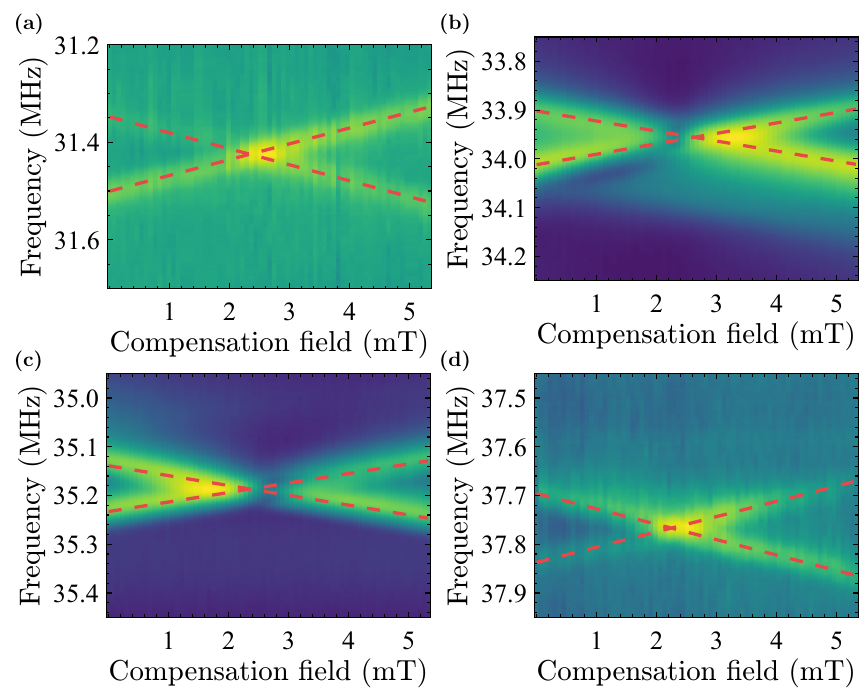}
    \caption{\textbf{Raman heterodyne scattering measurements.} The colormaps show the amplitdue of the RHS signal as a function of the RF excitation frequency and the applied compensation field along $b$ axis. The plots shows all the spin transitions between the $\ket{3_g}$, $\ket{4_g}$ and $\ket{5_g}$, $\ket{6_g}$ doublets, at frequencies $\omega_{4,5}$ (a), $\omega_{3,5}$ (b), $\omega_{4,6}$ (c) and $\omega_{3,6}$ (d), respectively. At zero compensation field, each transition shows two peaks due to the two magnetically inequivalent subsites. These merge at 2.35 mT, when the net magnetic field lies entirely in the $D_1$–$D_2$ plane.}
    \label{fig:RHS spectra}
\end{figure}

In Fig.\ref{fig:RHS spectra}, we see eight distinct RHS resonances with zero compensation field along the $b$ axis. When the compensation field is applied, each resonance varies linearly with the field and the sub-site split of each resonance is symmetric, which is expected for a small $b$-axis field component. Each pair of resonance lines were fitted to a linear curve, from which we extracted sub-site splitting rates of 65, 42, 40, and 63~kHz/mT for the transition frequencies $\omega_{4,5}$, $\omega_{3,5}$, $\omega_{4,6}$, and $\omega_{3,6}$, respectively. From the linear fits, the crossing point, i.e. the $D_1$-$D_2$ plane, can be accurately estimated. In this case, the RF transition frequencies were $\omega_{4,5} = 31.425$, $\omega_{3,5} = 33.956$, $\omega_{4,6} = 35.187$, and $\omega_{3,6} = 37.767$ MHz at the crossing point.

\subsection{Accurate magnetic field vector estimation}
\label{sec:field_estimation}

The magnetic field vector can be determined from the RHS data directly, if all four crossing point RF transition frequencies can be accurately determined. This is done by comparing the measured frequencies to those predicted by the ground-state spin Hamiltonian \cite{Longdell2006,ZambriniCruzeiro2018a}, using a non-linear least squares fitting. From the crossing-point frequencies given above, the fitted magnetic field vector is $\mathbf{B} = (-22.7,\ 249.2,\ 0.0)~\mathrm{mT}$ in the $D_1$,$D_2$,$b$ crystal reference frame, with uncertainties of $(5.6,\ 1.9,\ 0.7)~\mathrm{mT}$. If we express the vector in spherical angles $\mathbf{B} = \norm{\mathbf{B}}(\cos{\phi}\sin{\theta},\sin{\phi}\sin{\theta},\cos{\theta})$, we have $\norm{\mathbf{B}} = 250.3~\mathrm{mT}$, with angles $\phi = 95.2^\circ$ and $\theta = 90^\circ$, i.e. the estimated field is in the $D_1-D_2$ plane with a relative angle to the $D_2$ axis of $5.2^\circ$. The RMS value of the fit residuals for the RF frequencies was 12 kHz, demonstrating the high precision of this method for reconstructing both the magnitude and orientation of the magnetic field. Knowing the field vector, it is now possible to accurately predict all 36 optical-hyperfine transition frequencies, which is critical for applications such as quantum memories.

The field can also be estimated for zero compensation field applied to the $b$ axis, using all eight RF transitions of the two magnetic subsites. We find the field vector $\mathbf{B} = (-22.4,\ 248.9,\ 3.1)~\mathrm{mT}$, with similar RMS fit residuals and parameter uncertainties as for the cross point. By comparing the two field vectors, we see that the Helmholtz coils supplying the compensation field is well aligned in the $b$ axis.

Another method for estimating the B field vector is to employ SHB and to compare the hole and anti-hole frequencies to those predicted by the spin Hamiltonian. If the setup does not include the required RF coil for RHS, or if not enough RHS lines can be measured, due for instance to the frequency tuneability of the resonant circuit, then SHB is an alternative method. Also, from the SHB one gains information from both the ground and excited state, potentially making the field estimation more accurate, although for our field we did not observe any notable difference in accuracy.

For the SHB measurements, we do not apply any CC before the SHB, because in order to see a maximum number of holes and anti-holes, it is advantageous to work with as many frequency classes as possible. We burnt a single hole in the centre of the inhomogeneous broadening, and measured the absorption spectrum using a readout pulse scanned over 300 MHz. Here we present SHB measurements at the cross point, i.e. after having tuned the field into the $D_1-D_2$ plane as explained in Sec. \ref{sec:field_tuning_D1D2}.

\begin{figure}
    \centering
    \includegraphics[width=1\linewidth]{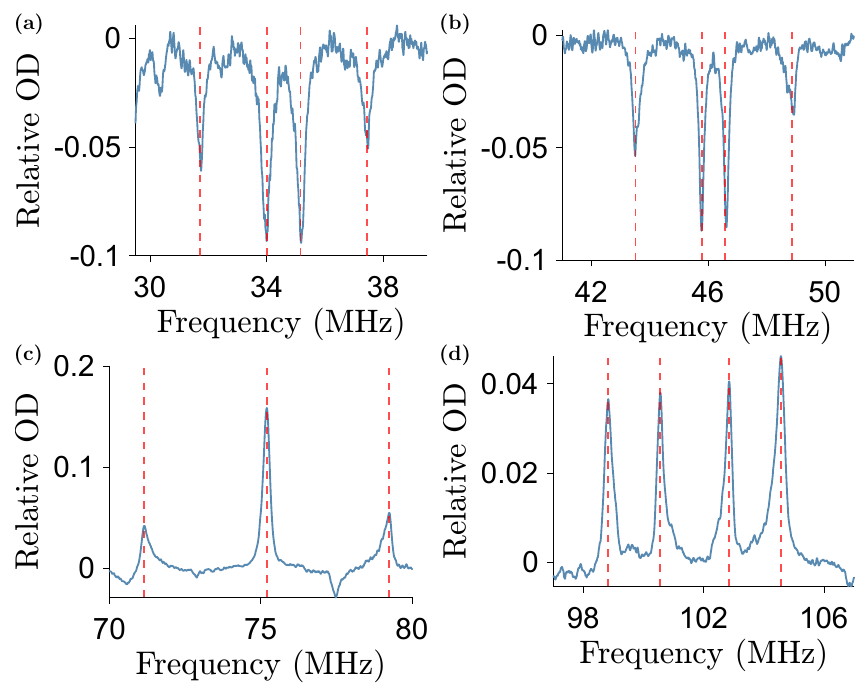}
    \caption{\textbf{Spectral hole burning measurements}. Eight well-resolved and strong spectral holes (lower panels) and anti-holes (upper panels) were measured for the field estimation. The holes give information on the excited state splittings, while these strong anti-holes give information on the ground state splittings. The red dashed lines indicate the transition frequencies calculated from the spin Hamiltonian model for the magnetic field vector estimated from these SHB measurements, see text for details, showing excellent agreement with the measured spectral features (deviations $<5$~kHz). Note that these measurements were done after tuning the field into the $D_1-D_2$ plane.}

    \label{fig:SHB spectra}
\end{figure}

As shown in Fig.~\ref{fig:SHB spectra}, eight spectral holes and eight anti-holes were selected based on their spectral isolation and signal-to-noise ratio. A detailed view of the optical depth spectrum reveals distinct anti-holes associated with specific ground-state hyperfine splittings. Fig.~\ref{fig:SHB spectra}(a) displays anti-holes corresponding to transitions between $\ket{3_g}$, $\ket{4_g}$ and $\ket{5_g}$, $\ket{6_g}$, centered around zero-field ground-state hyperfine splitting of 34.54~MHz. Similarly, Fig.~\ref{fig:SHB spectra}(b) shows anti-holes between $\ket{3_g}$, $\ket{4_g}$ and $\ket{1_g}$, $\ket{2_g}$, centered on 46.25~MHz. Fig.~\ref{fig:SHB spectra}(c) presents holes corresponding to transitions between $\ket{1_e}$, $\ket{2_e}$ and $\ket{3_e}$, $\ket{4_e}$, centered on the excited-state zero-field splitting of 75.03~MHz. Fig.~\ref{fig:SHB spectra}(d) displays holes associated with splittings between $\ket{3_e}$, $\ket{4_e}$ and $\ket{5_e}$, $\ket{6_e}$, centered on 101.65~MHz.

The measured SHB frequencies were compared to those calculated from the ground and excited state spin Hamiltonians \cite{ZambriniCruzeiro2018a}. For each magnetic field configuration, the model generates all 31 spectral holes and 930 anti-holes arising from 36 atomic classes. From this complete set, we identify the theoretical features corresponding to the experimentally observed transitions. A least-squares optimization is then performed to determine the magnetic field vector, giving the field vector $\mathbf{B} = (-26.9,\ 227.5,\ 0.0)~\mathrm{mT}$ with uncertainties $(0.3,\ 2.1,\ 0.1)~\mathrm{mT}$, i.e. $\norm{\mathbf{B}} = 229.1~\mathrm{mT}$ with angles $\phi = 96.7^\circ$ and $\theta = 90^\circ$ in spherical coordinates.

The field vectors estimated from the RHS and SHB measurements are different in magnitude and orientation, as these measurements were taken for two slightly different spacings of the permanent magnets inside the cryostat. The measurements presented here serve to showcase the methods and the precision with which the field can be estimated, using either RHS or SHB. They also provide a critical test of the spin Hamiltonians, which accurately predicts all measured spectral lines, with errors in the 5-12 kHz regime.

\subsection{Optical Rabi frequency measurements}
\label{sec:Rabi_measurements}

In this section, we describe the optical Rabi frequency measurements of 21 transitions out of the 36 possible optical-hyperfine transitions, which will be used in Sec. \ref{sec:branching_ratio_estimation} to estimate the $6 \times 6$ branching ratio table.

When a two-level system is driven coherently, the population oscillates between the states at the Rabi frequency

\begin{equation}
\Omega_{i,j} = \frac{\mu_{i_g,j_e}E}{\hbar},
\label{eq:Rabi_a}
\end{equation}

\noindent where $E$ is the electric field amplitude and $\mu_{i_g,j_e}$ the dipole matrix element between the ground state $\ket{i_g}$ and the excited state $\ket{j_e}$. As explained in Sec. \ref{sec:qm_with_field_B}, the dipole element $\mu_{i_g,j_e} = \mu_{g,e} \sqrt{\gamma_{i,j}}$, where $\gamma_{i,j}$ is the overlap between the nuclear spin states, i.e. the branching ratio element, and $\mu_{g,e}$ is the global transition dipole moment between the electronic ground and excited states, which is independent of the spin state. Eq. \ref{eq:Rabi_a} can be re-written as

\begin{equation}
\Omega_{i,j} = \Omega_{g,e} \sqrt{\gamma_{i,j}},
\label{eq:Rabi_b}
\end{equation}

\noindent where

\begin{equation}
\Omega_{g,e} = \mu_{g,e} E/ \hbar.
\label{eq:Rabi_c}
\end{equation}

\noindent The goal of Sec. \ref{sec:branching_ratio_estimation} is to estimate all 36 elements $\gamma_{i,j}$ and $\Omega_{g,e}$ through a restricted set of measurements. The dipole moment $\mu_{g,e}$ can then be calculated from the estimated $\Omega_{g,e}$ and the known field amplitude $E$.

To address an individual transition $\ket{i_g} \leftrightarrow \ket{j_e}$, we use CC and SP to polarize the ions from a specific frequency class into the ground state $\ket{i_g}$ over a frequency range of 1.5 MHz, using the methods explained in Sec. \ref{sec:simulation_opt_pump_num_model}. The six CC frequencies were adapted in order to address the specific transition, and we verified that the selected CC frequencies indeed produced a single frequency class of the transition we wanted to address, using the numerical simulator as described in Sec. \ref{sec:simulation_opt_pump_EuYSO_example}.

\begin{figure}[t]
    \centering
    \includegraphics[width=0.40\textwidth]{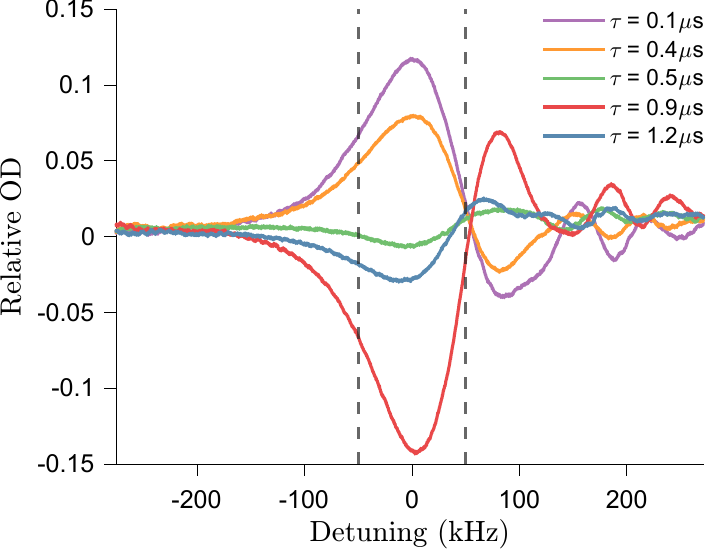}
    \caption{\textbf{Coherent excitation of a spectrally narrow ensemble.} The figure shows the narrow spectral peak created on the $\ket{6_g} \leftrightarrow \ket{5_e}$ transition. When driven coherently with a pulse of duration $\tau$, the optical depth (OD) of the peak oscillates at the Rabi frequency, showing both net absorption (positive OD) and amplification (negative OD). The dashed lines indicate the frequency range used for integration to quantify population transfer, cf. Fig \ref{fig:Rabi_three_oscillations}.}
    \label{fig:Rabi_example_peaks}
\end{figure}

To be able to produce high-contrast Rabi oscillations, one should reduce the inhomogeneous broadening of the driven ensemble of ions. A series of optical pulses pumped away most of the ions in the trench produced by CC and SP, but left behind a narrow peak of absorption on the $\ket{i_g} \leftrightarrow \ket{j_e}$ transition, in the middle of the otherwise transparent trench, with a width around 100 kHz, as shown in Fig. \ref{fig:Rabi_example_peaks}. See for instance Refs \cite{Nilsson2004,Rippe2005,Guillot-Noel2009} for the general approach. After the preparation step, the narrow peak is excited by a resonant pulse of varying duration $\tau$. The optical depth of the peak was measured for each pulse duration, using a readout pulse that was scanned over 500 kHz, see Fig. \ref{fig:Rabi_example_peaks}.

\begin{figure*}
    \centering
    \includegraphics[width=\textwidth]{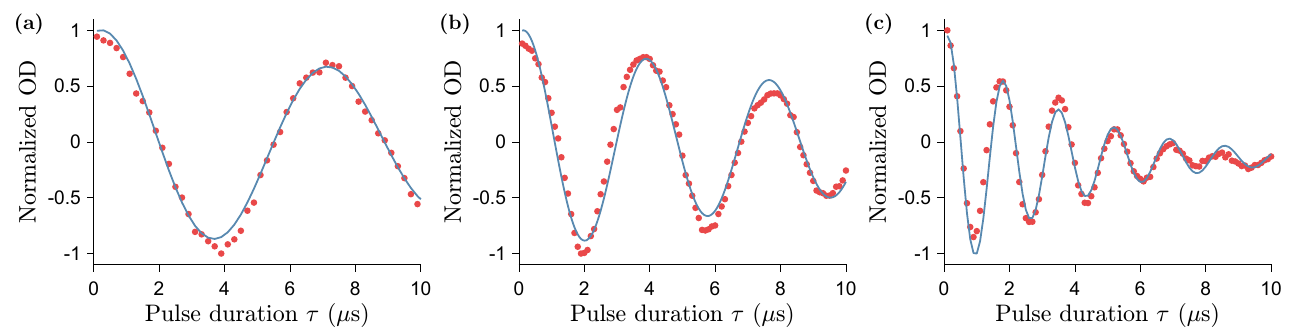}
    \caption{\textbf{Rabi oscillations}. Here we show three examples of Rabi oscillations and the fitted damped cosine functions, on transitions a) $\ket{5_g} \leftrightarrow \ket{3_e}$, b) $\ket{3_g} \leftrightarrow \ket{3_e}$ and c) $\ket{6_g} \leftrightarrow \ket{5_e}$, resulting in fitted Rabi frequencies of 145 $\pm $1 kHz,  266 $\pm $2 kHz and  581$\pm $6 kHz, respectively. Each experimental point represents the mean OD integrated between the dashed lines in Fig. \ref{fig:Rabi_example_peaks}, normalized to the range [-1,1].}   
    \label{fig:Rabi_three_oscillations}
\end{figure*}

In total, we were able to observe clear Rabi oscillations for 21 different transitions, with three examples shown in Fig. \ref{fig:Rabi_three_oscillations}. Each transition was driven with a different peak power in the excitation pulse, as the acousto-optic modulator that created all the optical pulses was working at a different RF frequency. To account for this, we power calibrate all the Rabi frequencies with respect to a reference transition. All 21 raw and power-calibrated Rabi frequencies are shown in Table \ref{tab:exp_Rabi_freqs}.

\begin{table}[b]
\centering
\caption{\textbf{Measured optical Rabi frequencies}. Calibrated data refers to the Rabi frequency normalized to a common reference laser power, here the $\ket{3_g} - \ket{6_e}$ transition, and the value in parentheses is the standard error from the fit. Raw Rabi frequency is the directly observed oscillation frequency at the actual peak power listed in the last column.}
\begin{tabular}{@{}l
                S[table-format=3.1(2)]
                S[table-format=3.0(1)]
                S[table-format=1.3]@{}}
\toprule
Transition & {Calibrated (kHz)} & {Raw (kHz)} & {Power (W)} \\
\midrule
$\ket{6_g} - \ket{3_e}$ & 175 \pm 1 & 147 \pm 1 & 0.276 \\
$\ket{6_g} - \ket{4_e}$ & 279 \pm 5 & 248 \pm 4 & 0.308 \\
$\ket{6_g} - \ket{5_e}$ & 604 \pm 6 & 581 \pm 6 & 0.360 \\
$\ket{5_g} - \ket{3_e}$ & 314 \pm 2 & 145 \pm 1 & 0.083 \\
$\ket{5_g} - \ket{5_e}$ & 164 \pm 10 & 132 \pm 8 & 0.252 \\
$\ket{5_g} - \ket{6_e}$ & 632 \pm 7 & 598 \pm 7 & 0.349 \\
$\ket{4_g} - \ket{2_e}$ & 202 \pm 3 & 120 \pm 2 & 0.138 \\
$\ket{4_g} - \ket{4_e}$ & 607 \pm 12 & 305 \pm 6 & 0.099 \\
$\ket{4_g} - \ket{5_e}$ & 308 \pm 5 & 303 \pm 5 & 0.378 \\
$\ket{3_g} - \ket{1_e}$ & 218 \pm 2 & 140 \pm 1 & 0.161 \\
$\ket{3_g} - \ket{3_e}$ & 591 \pm 4 & 266 \pm 2 & 0.079 \\
$\ket{3_g} - \ket{4_e}$ & 148 \pm 6 & 124 \pm 5 & 0.275 \\
$\ket{3_g} - \ket{6_e}$ & 313 \pm 2 & 313 \pm 2 & 0.390 \\
$\ket{2_g} - \ket{1_e}$ & 143 \pm 27 & 124 \pm 23 & 0.292 \\
$\ket{2_g} - \ket{2_e}$ & 695 \pm 7 & 212 \pm 2 & 0.036 \\
$\ket{2_g} - \ket{3_e}$ & 208 \pm 3 & 153 \pm 2 & 0.211 \\
$\ket{1_g} - \ket{1_e}$ & 681 \pm 7 & 208 \pm 2 & 0.036 \\
$\ket{1_g} - \ket{2_e}$ & 741 \pm 32 & 231 \pm 10 & 0.038 \\
$\ket{1_g} - \ket{3_e}$ & 123 \pm 4 & 89 \pm 3 & 0.203 \\
$\ket{1_g} - \ket{4_e}$ & 211 \pm 4 & 155 \pm 3 & 0.211 \\
$\ket{1_g} - \ket{6_e}$ & 139 \pm 7 & 125 \pm 6 & 0.317 \\
\bottomrule
\end{tabular}
\label{tab:exp_Rabi_freqs}
\end{table}

\section{Estimating the optical hyperfine branching ratios}
\label{sec:branching_ratio_estimation}

To quantify the relative oscillator strengths of optical transitions within the $6 \times 6$ hyperfine manifold of $^{151}\mathrm{Eu}^{3+}\!:\!\mathrm{Y}_2\mathrm{SiO}_5$, we reconstruct the branching ratio matrix $\gamma$ using the set of measured optical Rabi frequencies. There are 21 experimental data points reported in Table \ref{tab:exp_Rabi_freqs}, but there are 36 elements in total that need to be estimated in the  $\gamma$ matrix. However, the number of independent elements is reduced by two independent normalization constraints:

\begin{align}
    \sum_i \gamma_{ij} &= 1 \quad \text{for all ground states } |i_g\rangle, \\
    \sum_j \gamma_{ij} &= 1 \quad \text{for all excited states } |j_e\rangle.
\end{align}

\noindent These reflect conservation of total transition probability for each initial and final state. These double-normalization conditions impose $2 \times 6 = 12$ linear constraints on the 36 elements of $\gamma$, one of which is redundant due to the fixed total sum $\sum_{i,j} \gamma_{ij} = 6$, resulting in 11 constraints. Hence, 25 independent parameters of $\gamma$ need to be determined, which is still a larger number than the 21 measurements.

To solve this problem even in principle, it helps to think of it as a system of linear equations, expressed in a standard matrix form. For a solvable system in principle, we need at least 36 linear equations, i.e. the matrix should be sqaure and invertable. The 11 constraints above fill 11 rows of the matrix of linear equations, while the 25 additional rows are the measurements results we supply. However, any arbitrary choice of the 25 measurements will not yield an invertable system of equations. Hence, we should check that, for the 25 elements we select as independent parameters in the $\gamma$ matrix, the system of equations is in principle invertable. Our extensive tests indicate that a non-linear fitting algorithm will not converge unless the elements are chosen such that the system is in principle invertable.

\begin{table*}
\centering
\renewcommand{\arraystretch}{1.4}
\caption{
Fitted branching ratio matrix $\gamma$ (in \%) obtained from global least-squares fitting.
The fit yields a maximum Rabi frequency of $740 \pm 10$~kHz. Uncertainties (in parentheses) correspond to one standard deviation. The 21 bold elements indicate the transitions that were directly measured and used as fixed constraints in the fit.}
\label{tab:gamma_fit}
\begin{tabular}{rrrrrrr}
\toprule
 & $\ket{1_e} $ & $\ket{2_e}$ & $\ket{3_e}$ & $\ket{4_e}$ & $\ket{5_e}$ & $\ket{6_e}$ \\
\midrule
$\ket{6_g}$ & 0.0        & 0.0        & \textbf{6.0(1.2)}  & \textbf{15.5(1.8)} & \textbf{73.0(2.2)} & 5.6 \\
$\ket{5_g}$ & 0.0        & 0.0        & \textbf{17.2(1.9)} & 5.35               & \textbf{4.8(1.1)}  & \textbf{72.7(2.3)} \\
$\ket{4_g}$ & 2.54       & \textbf{7.2(1.4)}  & 3.81               & \textbf{64.8(2.4)} & \textbf{16.8(2.1)} & 4.93 \\
$\ket{3_g}$ & \textbf{9.0(1.6)}  & 3.61               & \textbf{62.9(2.5)} & \textbf{4.0(1.0)}  & 2.11               & \textbf{18.5(2.3)} \\
$\ket{2_g}$ & \textbf{3.8(1.0)}  & \textbf{85.5(3.8)} & \textbf{7.5(1.4)}  & 2.7                & 3.4                & 0.0 \\
$\ket{1_g}$ & \textbf{84.7(5.4)} & \textbf{3.6(1.0)}  & \textbf{2.7(0.9)}  & \textbf{7.8(1.4)}  & 0.0                & \textbf{3.5(1.0)} \\
\bottomrule
\end{tabular}
\end{table*}

Given our insufficient number of data points, we need to make an assumption of the matrix elements for 4 additional elements, and check that the system of equations is in principle invertable for that choice. To this end, we consider the theoretical $\gamma$ matrix, shown as Table \ref{tab:gamma_theory} in the Appendix, calculated for the field magnitude and orientation of the Rabi experiment. Some elements are very weak, particularly the four elements of the transitions between the $\ket{5_g},\ket{6_g}$ and $\ket{1_e},\ket{2_e}$ doublets. These elements we do not expect to be able to measure with our maximum peak power, hence we set their corresponding experimental value as zero in the matrix. With these additional elements, we checked that the system of equations is indeed invertable for our specific 25 selected elements.

In the experiment, we do not directly measure the elements of $\gamma$, but the associated Rabi frequencies. We therefore also need to fit a 26th parameter, which is the global optical Rabi frequency $\Omega_{g,e}$ in Eq. \ref{eq:Rabi_b}. The fitting is done with a nonlinear least-squares optimization, 
subject to positivity constraints ($0 < \gamma_{ij} < 1$), with initial values for the fitted 25 elements $\gamma_{ij}$ taken from theoretical estimates derived from the spin Hamiltonians.

The optimization converged reliably, yielding the fitted branching ratio shown in Table \ref{tab:gamma_fit}. The reconstructed matrix adheres to the required row and column sum constraints, with deviations below 5.2\%. The fitted branching ratio table globally agrees well with the theoretical table. This comparison provides a severe test for the spin Hamiltonians for both the ground and excited state, and the relative orientations of their tensors \cite{Longdell2006,ZambriniCruzeiro2018a}. It should be emphasized that the Hamiltonians were measured for fields at least one order of magnitude weaker, and that no experimental information existed, nor were any comparison made, with respect to the branching table.

The fitted global optical Rabi frequency is $\Omega_{g,e} = 740 \pm 10 ~\mathrm{kHz}$. To calculate the optical dipole moment $\mu_{g,e}$ in Eq. \ref{eq:Rabi_c}, we need to estimate the electric field amplitude. The field strength is related to the laser power $P$ and beam area $A$ through $E =\sqrt{ 2 P/(A n \epsilon_0 c)}$, where $n = 1.8$ is the index of refraction of the \yso{} crystal, $\epsilon_0$ the vacuum permittivity and $c$ the speed of light in vacuum. For the effective beam area, we take the choice $A = \pi w_0^2$ \cite{Lauritzen2012}, where $w_0$ is the normal beam waist radius. For these measurements, with a power $P = 390~\mathrm{mW}$ and waist $w_0 = 102~\mu\mathrm{m}$, the field is then $E = 70.6$ kV/m, resulting in an electric dipole moment $\mu_{g,e} = (6.94 \pm 0.09) \times 10^{-33}$  $\mathrm{C} \cdot \mathrm{m}$ for the \eutransition{} transition in \euyso{}. This value is in reasonable agreement with the previously reported value of $5 \times 10^{-33}~\mathrm{C} \cdot \mathrm{m}$~\cite{Lauritzen2012} at zero magnetic field.

\section{Discussion and outlook}
\label{sec:discussion}

In this work, we have described simulation tools and methodologies for performing experiments and detailed spectroscopic measurements of specific optical-hyperfine transitions in \euyso{} under applied magnetic field. Generally, we have found that the published spin Hamiltonians \cite{Longdell2006,ZambriniCruzeiro2018a,Ma2018} accurately describe the energy spectrum and relative transition strengths under magnetic field, which are key tools for designing current and future experiments in \euyso{}. Specifically, these tools were crucial for the recent work in Ref. \cite{Chen2025}, where we demonstrated efficient and reversible optical and spin manipulation in an AFC spin-wave optical memory.

Furthermore, we experimentally measured optical Rabi frequencies of a large number of optical-hyperfine transitions, from which we could estimate the 6x6 branching ratio matrix of relative transitions strengths, as well as the optical dipole moment of the \eutransition{} transition in \euyso{}. This matrix is particularly sensitive to the accuracy of the spin Hamiltonians. Indeed, as discussed in Ref. \cite{ZambriniCruzeiro2018a}, the branching ratio matrix is sensitive to the signs and relative angles of the Zeeman and quadrupole tensors of the spin Hamiltonian. It was concluded that those are difficult to determine based solely on the measured energy spectrum, and several possible solutions for the tensors were considered in Ref. \cite{ZambriniCruzeiro2018a}. As these solutions could not be differentiated by the energy spectrum measurements, the solution was selected by comparing the theoretically predicted and experimentally measured 3x3 branching ratio table without magnetic field. Our analysis of the complete 6x6 branching matrix under field shows that the solution proposed in Ref. \cite{ZambriniCruzeiro2018a} is indeed the correct one, as the other solutions generally do not agree as well with our experimentally estimated branching ratio matrix. Our results indicate that the spin Hamiltonians can be used with confidence when designing future experiments in \euyso{} under applied magnetic fields.

\begin{figure*}[ht]
    \centering
    \includegraphics[width=\linewidth]{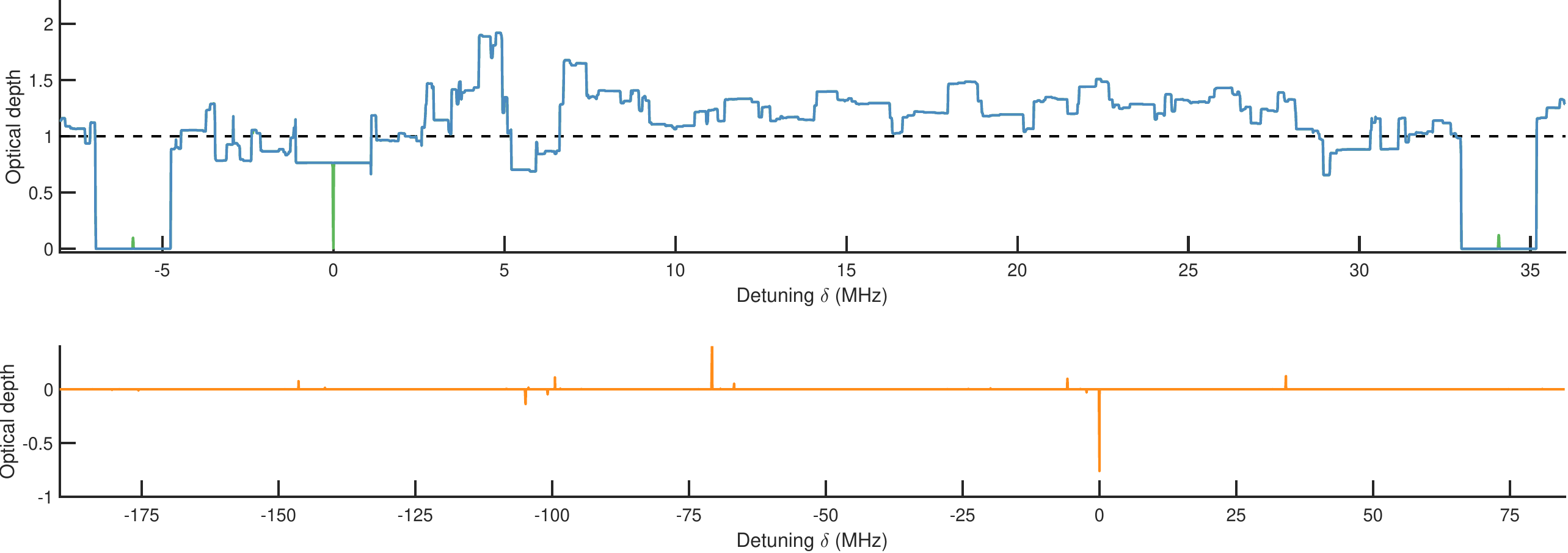}
    \caption{\textbf{Numerical simulation of the spectrum with true branching ratio values.} Simulated absorption spectra using the same frequencies and parameters as in the simulations shown in Fig. \ref{fig:spectrum_simulation}, except for the branching ratio which were here computed from the spin Hamiltonians for the field of 231 mT along the $D_2$ axis.}
    \label{fig:spectrum_simulation_D2_dm}
\end{figure*}

\acknowledgments 

This research was funded by the Swiss State Secretariat for Education, Research and Innovation
(SERI) under contract number UeM019-3.

\appendix
\section{Optical pumping simulations with true branching ratio table}
\label{app:simulation_true_dm}

In Fig. \ref{fig:spectrum_simulation_D2_dm}, we show the same simulation as in Fig. \ref{fig:spectrum_simulation} in Sec. \ref{sec:qm_with_field_B}, but with branching ratio elements $\gamma_{i,j}$ computed with the spin Hamiltonians, cf. Table \ref{tab:gamma_theory}.

Comparing Figs \ref{fig:spectrum_simulation} and \ref{fig:spectrum_simulation_D2_dm}, one sees that there is significantly more absorption in the zero-frequency trench, i.e. on the \fone{} transitions for this single frequency class, after the spin polarisation step when using the true branching ratios. This is due to the fact the \fone{} transition is a strong transition, having branching element of $\gamma_{5,6} = 0.764$ according to the spin Hamiltonians. Recall that in Fig. \ref{fig:spectrum_simulation}, all $\gamma_{i,j}$ elements were set to 1/6.

To understand the resulting absorption depth after the class cleaning and spin polarisation steps, we should examine Eq. \ref{eq:theory_spectrum} more carefully. We choose to specifically look at the centre of the inhomogeneous broadening, i.e. at the zero frequency $\delta = 0$, and we only count the resonant contribution to the absorption at this wavelength. Hence we count the $i \times j = 36$ contributing frequency classes resonating at $T_{i,j}(\delta_{g,e}) = 0$, for  $i \times j$ different $\delta_{g,e}$ detunings within the inhomogeneous broadening. That is, we only count the contributions at resonance $H(\delta - T_{i,j}(\delta_{g,e}) = 0)$. Furthermore, we assume that each class contributes equally to the broadening, in the sense that we assume equal $G(\delta_{g,e})$ for these detunings $\delta_{g,e}$. The resonant contribution to the $\delta = 0$ absorption is then simply

\begin{equation}
    f(\delta) = \kappa \sum_{i,j} \gamma_{i,j} \rho_{i}
    \label{eq:theory_spectrum_res_contr}
\end{equation}

\noindent where $\kappa$ is new normlization factor that is independent of $\gamma_{i,j}$ and $\rho_{i}(\delta_{g,e})$. Here we have supressed the sum over $\delta_{g,e}$, it is implicit that when changing indeces $i$,$j$ in the sum, we need to find the corresponding detunings $\delta_{g,e}$ such that we have a resonant contribution $\delta - T_{i,j}(\delta_{g,e}) = 0$. 

The effect of the class cleaning and the spin polarisation can now be understood by Eq. \ref{eq:theory_spectrum_res_contr}. Without any optical pumping, all $i \times j = 36$ classes contribute an equal amount of population, as we assume an equal thermal distribution over the hyperfine states, i.e. $\rho_i(\delta_{g,e})=1/6$. The equation then becomes

\begin{equation}
    f(\delta) = \kappa \frac{1}{6} \sum_{i,j} \gamma_{i,j} = \kappa \frac{1}{6} \times 6 = \kappa,
\end{equation}

\noindent given the definition of the $\gamma_{i,j}$ elements. In the case of class cleaning, ideally only a single frequency class $\delta_{g,e}$ as non-zero population elements. In the simulations of Figs \ref{fig:spectrum_simulation} and \ref{fig:spectrum_simulation_D2_dm}, the choosen class cleaning frequencies selects the frequency class for which the \fone{} transition resonates at $\delta = 0$, and the spin polarisation step pumps all the population of this class into the $\ket{5_g}$ state, i.e. $\rho_{5} = 1$. The other 35 frequency classes resonantly contributing to the $\delta = 0$ absorption has $\rho_{i} = 0$, for $i \neq 5$, assuming perfect class cleaning. Hence, there is only a single non-zero contribution in the sum of Eq. \ref{eq:theory_spectrum_res_contr}:

\begin{equation}
    f(\delta) = \kappa \gamma_{5,6} \rho_{5} = \kappa \gamma_{5,6}
\end{equation}

\noindent The conclusion is that a fully class cleaned and spin polarised ensemble will have an optical depth of $\gamma_{i,j}$ times the initial, unpumped optical depth, where $\gamma_{i,j}$ is the branching element of the relevant transition. This conclusion holds if we assume equal thermal population for the unpumped spectrum, and a sufficiently wide inhomogeneous broadening with respect to the frequency range spanned by all the $T_{i,j}$ transitions of each class. Practically, this means that only working with one frequency class out of the 36 possible ones does not significantly reduce the available absorption, provided that it can be efficiently spin polarised into a single hyperfine state through optical pumping, which is not completely intuitive when first considering the problem.

\section{Theoretical branching ratio table}
\label{app:theory_table}
Table~\ref{tab:gamma_theory} shows the theoretical branching ratio table computed from the spin Hamiltonians, as discussed in Sec. \ref{sec:qm_with_field_B}. It is computed for the magnetic field used in the optical Rabi frequency measurements in Sec. \ref{sec:Rabi_measurements}, i.e. $\mathbf{B} = (-30.8,\ 227.0,\ 0.0)~\mathrm{mT}$ in the $D_1$,$D_2$,$b$ crystal reference frame, with uncertainties $(1.4,\ 0.4,\ 0.1)~\mathrm{mT}$.

\begin{table*}[t]
\centering
\renewcommand{\arraystretch}{1.4}
\caption{Theoretical branching ratio matrix $\gamma_{\mathrm{theory}}$ (in \%) calculated from spin Hamiltonian model, see Appendix \ref{app:theory_table}.}
\label{tab:gamma_theory}
\begin{tabular}{rrrrrrr}
\toprule
 & $\ket{1_e} $ & $\ket{2_e}$ & $\ket{3_e}$ & $\ket{4_e}$ & $\ket{5_e}$ & $\ket{6_e}$ \\
\midrule
$\ket{6_g}$ & 0.8  & 0.6  & 4.4  & 14.0 & 75.4 & 4.5 \\
$\ket{5_g}$ & 0.7  & 1.1  & 14.2 & 4.3  & 4.7  & 75.0 \\
$\ket{4_g}$ & 2.0  & 8.6  & 4.1  & 68.6 & 16.3 & 0.4 \\
$\ket{3_g}$ & 10.3 & 2.2  & 66.4 & 4.0  & 0.4  & 16.7 \\
$\ket{2_g}$ & 3.4  & 83.6 & 7.6  & 2.3  & 2.8  & 0.2 \\
$\ket{1_g}$ & 82.9 & 3.8  & 3.2  & 6.8  & 0.2  & 3.0 \\
\bottomrule
\end{tabular}
\end{table*}

\bibliography{qmcommon}

\end{document}